\documentclass[aps,prd,groupedaddress]{revtex4}
\usepackage{graphicx,epsf,amssymb,amsmath,latexsym}
\usepackage{epsfig}
\def\ben{\begin{equation}}
\def\een{\end{equation}}
\def\bena{\begin{eqnarray}}
\def\eena{\end{eqnarray}}

\begin{document}

\title{Gravitational Trapping Near Domain Walls and Stable Solitons}

\vspace{.3in}

\author{Eduardo Guendelman}
\email{guendel@bgu.ac.il}
\author{Idan Shilon}
\email{silon@bgu.ac.il}

\affiliation{Physics Department, Ben-Gurion University of the Negev, Beer-Sheva 84105, Israel }

\vskip.3in

\begin{abstract}

In this work, the behavior of test particles near a domain wall of
a stable false vacuum bubble is studied. It is shown that matter
is naturally trapped in the vicinity of a static domain wall, and
also, that there is a discontinuity in the test particle's
velocity when crossing the domain wall. The latter is unexpected
as it stands in contrast to Newtonian theory, where infinite
forces are not allowed. The weak field limit is defined in order
to show that there is no conflict with the non-relativistic
behavior of gravitational fields and particle motions under these
conditions.
\end{abstract}

\maketitle

\setcounter{equation}{0}
\section{Introduction}

The formation of domain walls can be caused, for example, by a spontaneous symmetry breaking of a discrete
symmetry at a phase transition. These two dimensional topological defects are likely to occur in the very
early universe, where the universe has cooled down through some critical temperature and the scalar field
dominating the universe acquired a non-zero value. Therefore, domain walls may be significant for the
evolution of the universe \cite{walls}.

In general, domain walls are forming when the spacetime has two or more disconnected regions. Our
set up is made out of a spherically symmetric domain wall localized at $r=R(\tau)$, with a false vacuum region in the interior ($r<R$) and a true vacuum region in the exterior ($r>R$). One should note that the domain wall location is, in general, a function of the time (one can get a dynamical wall solution if one uses a time dependent scalar field). We will work in the thin-wall approximation, which assumes that the thickness of the wall is infinitely small compared to all other length scales in the problem. Therefore, put more visually, we are considering a bubble of false vacuum.

Several different models, which use the physical system we have presented here, have been suggested. Many of these models are considering the creation of a "baby universe" from a false vacuum bubble which detaches, classically or via tunneling, from the original spacetime as it goes through an inflationary phase \cite{blau},\cite{fgg}. Other models consider the possibility for a stable, elementary particle like, bubble using additional matter terms in the energy density of the domain wall or a negative surface tension (or both) \cite{port},\cite{stable}.

In this work we shall study the motion of test particles in the vicinity of a stable false vacuum bubble (i.e. a stationary domain wall). As an example of how to obtain a stable false vacuum bubble, we will review the work of Guendelman and Portnoy \cite{port}: One can stabilize a false vacuum bubble by introducing gauge fields that live in the 2+1 dimensional spacetime of the domain wall, together with a surface tension. The gauge fields define a 2+1 dimensional gauge theory on the surface of the brane. This leads to an additional term in the effective surface tension of the domain wall, which gives rise to the solution of a stable bubble.

The structure of the paper is as follows: In the next section we will review the metric matching conditions on the brane using the Gauss-Codacci formalism to write Einstein's equations. In Sec. III we explain the way to obtain a stationary, stable, bubble configuration and in Sec. IV we derive the geodesic equation of motion for test particles in the vicinity of the brane. In Sec. V we discuss our results.

\section{The Matching Conditions}
\label{The Matching Conditions}

Our system consists of a domain wall which splits the spacetime into two regions, for each of which,
Einstein's equations are assumed to be satisfied separately. The geometric property of the system manifests
itself in the way that the domain wall is embedded in the two regions. In order to compare the two geometries we use the extrinsic curvature of the domain wall, induced by each of the two regions. The jump between the two extrinsic curvature tensors on the brane will yield the equation of motion of the domain wall. This is done using the Gauss-Codacci formalism, which is a method of viewing the four-dimensional spacetime as being sliced up into three-dimensional hypersurfaces \cite{mtw}. The resulting equation is Israel's junction conditions \cite{isr}:

\begin{eqnarray}
    \gamma^{i}_{~j} = -8\pi G(S^{i}_{~j} - \frac{1}{2}\delta^{i}_{~j}\mbox{Tr}S),
\end{eqnarray}

Where

\begin{eqnarray}
    \gamma_{ij} = \lim_{\epsilon\rightarrow 0}[K_{ij}(\eta =+\epsilon) - K_{ij}(\eta = -\epsilon)],
\end{eqnarray}

,$S_{\mu\nu}$ is the surface stress-energy tensor and $\eta$ is the normal coordinate to the brane. The energy-momentum tensor for the system under consideration can be written as

\begin{eqnarray}
    T_{\mu\nu} = \begin{cases} -\rho_{0}g_{\mu\nu},& \mbox{in the false vacuum region},\cr 0,& \mbox{in the true vacuum region,}\end{cases}
\end{eqnarray}

where $\rho_{0}$ is the false vacuum energy density. In the thin wall approximation, $T_{\mu\nu}$ has a
$\delta$-function singularity on the domain wall. Thus, one can define the surface stress-energy tensor by
writing

\begin{eqnarray}
    T_{\mu\nu} = S_{\mu\nu}\delta(\eta) + \mbox{regular terms}.
\end{eqnarray}

The metrics we use will be of the form

\begin{eqnarray}
    ds_{\pm}^2 = -A_{\pm}(r)dt_{\pm}^2 + A^{-1}_{\pm}(r)dr_{\pm}^2 + r^2d\Omega^2,
\end{eqnarray}

where the $+$ and $-$ are indices indicating the outside and inside regions, respectively. These spherically symmetric metrics satisfy the conditions \cite{avi}:

\begin{eqnarray}
    T^{t}_{~t} = T^{r}_{~r}~~\mbox{and}~~T^{\theta}_{~\theta} = T^{\phi}_{~\phi}.
\end{eqnarray}

A wide range of metrics can be written in this form, including, of course, the de-Sitter metric, The
Schwarzschild metric, The Reissner-Nordstrom metric, etc. The induced metric on the brane will be well defined if the internal and external radii coincide on the brane, which is obtainable from the demand that total area of the brane, as measured from both regions at the same proper time of the brane, will yield the same result and if the time flow on either side of the domain wall satisfies

\begin{eqnarray}
   (-A_{+}(R)\dot{t}_{+}^2+A_{+}^{-1}(R)\dot{R}^2)d\tau^2 = (-A_{-}(R)\dot{t}_{-}^2+A_{-}^{-1}(R)\dot{R}^2)d\tau^2.
\end{eqnarray}

The spherical symmetry of our system ensures us that the off-diagonal components of the extrinsic curvature
tensor vanish and that the angular components are related by $K_{\phi\phi} = \mbox{sin}^{2}\theta
K_{\theta\theta}$. Hence, one can conclude that the junction conditions are completely determined by the
$\theta\theta$ and $\tau\tau$ components of equation (1). The angular components yield the following equation of motion for a general domain wall \cite{blau}:

\begin{eqnarray}
    \beta_{-} - \beta_{+} = 4\pi G\sigma R,
\end{eqnarray}

where $\sigma$ is the energy density on the domain wall, $R$ denotes the value of the radial coordinate at
the wall (for both regions) and the $\beta_{\pm}$ are defined as

\begin{eqnarray}
    \beta_{+} = \pm (A_{+}(R) + \dot{R})^{1/2}~~\mbox{and}~~
    \beta_{-} = \pm (A_{-}(R) + \dot{R})^{1/2},
\end{eqnarray}

where the signs are to be determined by the geometric analysis of the problem. The proper-time component of
(1) yields just the proper-time derivative of equation (8). Thus, the latter determines all the properties of the solution to the general problem of our system. One should note that for geometries that do not contain wormholes, as the one we will consider in section IV, the signs of both $\beta_{+}$ and $\beta_{-}$ must be positive.

\section{Static and Stable Domain Walls}
\label{sec:StaticStableDomainWalls}

In general, the domain wall of the bubble will not be static. For
that reason, we wish to review and explain here the idea behind
the possibility of obtaining a stationary domain wall. Attempts to
describe a stable elementary particle with finite size has began a
long time ago (see for example the paper by Einstein from 1919
\cite{stable}, where he proposed to describe an elementary
particle as a bubble with an internal cosmological constant). The
first brane model was suggested by Dirac in 1962 \cite{stable},
where he thought of the electron as a charged conducting surface
with a positive surface tension which balances the repulsive
forces of the charge. In Dirac's paper gravity was not considered,
however. Yet, by using the Israel junction conditions one can
solve the corresponding problem with the presence of gravity.

The equation of motion for the domain wall is equivalent to the
equation of motion of a particle moving in one dimension with an
effective potential \cite{blau}. Naturally, the dynamic coordinate
for both equations is the bubble trajectory. If the surface
tension of the domain wall is some constant, the effective
potential will not have minima, and, therefore, one can not have a
solution for a static and stable bubble. In order to obtain a
local minimum in the effective potential, one should consider the
more general case, where the surface tension of the domain wall is
a function of the dynamic coordinate (i.e. $\sigma = \sigma(r)$).
The radial dependence of the effective surface tension yields the
possibility of obtaining a minimum value of the potential and,
thus, a stable configuration.

There are many examples for ways to obtain a stable false vacuum bubble. We choose to review the rather simple example of the model suggested by Guendelman and Portnoy (GP) \cite{port}. In the GP model, one views an elementary particle as a 2+1 dimensional brane embedded in a 3+1 bulk . Next, a 2+1 gauge theory is introduced on the brane surface. The action for the brane now takes the form

\begin{eqnarray}
    S = \sigma_{0}\int \sqrt{-h}d^{3}y + \lambda\int\sqrt{-h}F_{ab}F^{ab}d^{3}y,
\end{eqnarray}

where $a$ and $b$ take the values 0,1 and 2, $h = \mbox{det}(h_{ab})$ and $h_{ab}$ is the induced metric on
the brane. For the spherically symmetric bubble, the simplest non-trivial potential that respects the
spherical symmetry (up to a gauge transformation) is the magnetic monopole configuration

\begin{eqnarray}
    A_{\phi} = f(1-\mbox{cos}\theta),
\end{eqnarray}

which implies that $F_{\theta\phi} = f\mbox{sin}\theta$. The most general two dimensional spherically symmetric metric is given by

\begin{eqnarray}
    ds^2 = h_{ab}dy^ady^b = -d\tau^2 + r^2(\tau)d\Omega^2.
\end{eqnarray}

Therefore, we have $F_{ab}F^{ab} = 2f^2/r^4$, which means that the action can now be written as

\begin{eqnarray}
    S = 4\pi\sigma_{0}\int r^2(\tau)d\tau + 8\pi\lambda\int\frac{f^2}{r^2(\tau)}d\tau = 4\pi\int\left(\sigma_{0} + \frac{2\lambda f^2}{r^4(\tau)}\right)r^2(\tau)d\tau.
\end{eqnarray}

Hence, one can write as an effective surface tension:

\begin{eqnarray}
    \sigma(r) = \sigma_{0} + \frac{2\lambda f^2}{r^4} = \sigma_{0} + \frac{\sigma_{1}}{r^4},
\end{eqnarray}

with $\sigma_{1} = 2\lambda f^2$. If one calculates the effective potential from equation (8) using the
effective surface tension, $\sigma(r)$, one can obtain a local minimum of the potential (for $\lambda>0$).
This allows a stable configuration for the false vacuum bubble.

As a further example, which is also of interest in string theory, we mention the model of Alberghi, Lowe and Trodden \cite{stable}, which allows for a stable false vacuum bubble. This model considers a de-Sitter interior, a charged domain wall and a Reissner-Nordstrom exterior. A particular solution is where the surface tension force and pressure from the outside are cancelled by the electric force.

\section{Geodesic Motion Near the Domain Wall}
\label{sec:GeodesicMotionNearTheDomainWall}

In order to investigate the behavior of matter near a domain wall, we calculate the geodesic lines in the
vicinity of the domain wall. This will show how a test particle will "free fall" near the brane. Let us consider the proper time action functional:

\begin{eqnarray}
    S = \int\sqrt{-g_{\mu\nu}\frac{dx^{\mu}}{d\lambda}\frac{dx^{\nu}}{d\lambda}}d\lambda.
\end{eqnarray}

Applying the variational calculus to the latter will yield the geodesic equation. Thus, this action will be
used to describe the motion of test particles in the gravitational field of a static domain wall.

In order to analyze the problem in a proper way, we need to define a coordinate system which will be continuous across the domain wall. To that end, we need to redefine the radial coordinate, as well as the time flow on either side of the brane, so that the metric components will acquire equal values on the brane.

Let us begin with the radial coordinate. We define the new coordinate, $\eta$, as

\begin{eqnarray}
    \eta(r) = \int_{R}^{r}\frac{dr}{\sqrt{A(r)}}.
\end{eqnarray}

As a consequence, $\eta$ is negative in the interior, positive in the exterior and the brane is located at $\eta = 0$. Also, we have $g_{\eta\eta} = 1$ all over space.

The next step in making the metric continuous on the brane, is to
make sure that time flows in equal manner in both regions and that
both observers measure the same time on the brane. To that end, we
rescale the time coordinate of the interior region to be

\begin{eqnarray}
    dt_{-}^2 = \lambda^2 dt_{+}^{2},
\end{eqnarray}

where

\begin{equation}
    \lambda^2 = \frac{A_{+}(r=R)}{A_{-}(r=R)}.
\end{equation}

From condition (8), we see that for a static brane we have $0<\lambda^2<1$. Now, we can write the metrics as

\begin{eqnarray}
    ds_{+}^2 = -A_{+}(r)dt^2 + d\eta^2 + r(\eta)^2d\Omega^2,
\end{eqnarray}

in the exterior ($r>R$) and

\begin{eqnarray}
    ds_{-}^2 = -\lambda^2 A_{-}(r)dt^2 + d\eta^2 + r(\eta)^2d\Omega^2,
\end{eqnarray}

in the interior ($r<R$). We have omitted the $+$ index on the time coordinate since both metrics are referring to the same time coordinate now.

The domain wall is taken to be static, therefore, one concludes that the metrics are stationary. From the time independence of the metrics and the spherical symmetry of the system, Noether's theorem leads to

\begin{eqnarray}
    p_{0} = E = g_{00}\frac{dt}{d\tau} = \mbox{const.}
\end{eqnarray}

and also

\begin{eqnarray}
    p_{\phi} = l = g_{\phi\phi}\frac{d\phi}{d\tau} = r^{2}\frac{d\phi}{d\tau} = \mbox{const.}
\end{eqnarray}

Using

\begin{eqnarray}
    \frac{ds^2}{d\tau^2} = -1 = g_{\mu\nu}\frac{dx^{\mu}}{d\tau}\frac{dx^{\nu}}{d\tau} = g_{00}\left(\frac{dt}{d\tau}\right)^2 + \left(\frac{d\eta}{d\tau}\right)^2 + r^{2}\left(\frac{d\phi}{d\tau}\right)^2,
\end{eqnarray}

and also $d\eta^2 = \frac{1}{A}dr^2$, we can write

\begin{eqnarray}
    -1 = -g_{00}\left(\frac{dt}{d\tau}\right)^2 + \frac{1}{A}\left(\frac{dr}{d\tau}\right)^2 +
    r^{2}\left(\frac{d\phi}{d\tau}\right)^2.
\end{eqnarray}

Hence, we obtain the following equations of motion for a test
particle in both regions:

\begin{eqnarray}
    \mbox{for}~~ r<R:~ \frac{1}{2}\left(\frac{A_{+}(R)}{A_{-}(R)}\right)\dot{r}^2 + \frac{A_{+}(R)}{2A_{-}(R)}A_{-}(r)\left(1 + \frac{l^2}{r^2}\right) = \frac{1}{2}E^2,
\end{eqnarray}

\begin{eqnarray}
    \mbox{for}~~ r>R:~ \frac{1}{2}\dot{r}^2 + \frac{1}{2}A_{+}(r)\left(1 + \frac{l^2}{r^2}\right) = \frac{1}{2}E^2.
\end{eqnarray}

Both eqautions can be put in the form

\begin{eqnarray}
    \frac{1}{2}m_{eff}\dot{r}^2 + V_{eff} = E_{eff},
\end{eqnarray}

with the quantities

\begin{eqnarray}
    m_{eff} = \begin{cases} \frac{A_{+}(R)}{A_{-}(R)}, & ~ r < R, \cr 1,& ~ r>R, \end{cases}
\end{eqnarray}

\begin{eqnarray}
    V_{eff} = \begin{cases} \frac{1}{2}\frac{A_{+}(R)}{A_{-}(R)}A_{-}(r)\left(1 + \frac{l^2}{r^2}\right), & ~ r < R, \cr \frac{1}{2}A_{+}(r)\left(1 + \frac{l^2}{r^2}\right),& ~ r>R, \end{cases}
\end{eqnarray}

\begin{eqnarray}
    E_{eff} = \frac{1}{2}E^2.
\end{eqnarray}

We immediately see that we have obtained a continuous potential, but, as a consequence, a test particle crossing the brane will have a \textit{different effective mass} in each region.
Effectively, this corresponds to a discontinuity in the velocity
of the test particle, a discontinuity which indicates that an
infinite force (if we speak in Newtonian terms) have acted on the
test particle. If we evaluate the test particle's velocity at the
brane we see that in the exterior we have

\begin{equation}
    \dot{r}_{+}^2 = E^2 - A_{+}(R)\left(1 + \frac{l^2}{R^2}\right) = v^2,
\end{equation}

while in the interior we have

\begin{equation}
    \lambda^2\dot{r}_{-}^2 = E^2 - \frac{A_{+}(R)}{A_{-}(R)}A_{-}(R)\left(1 + \frac{l^2}{R^2}\right) = E^2 - A_{+}(R)\left(1 + \frac{l^2}{R^2}\right) = v^2.
\end{equation}

Thus, the jump in the magnitude of the velocity will be

\begin{equation}
 \Delta(\dot{r}^2) = \dot{r}_{-}^2 - \dot{r}_{+}^2 = v^2\left(\frac{1 - \lambda^2}{\lambda^2}\right),
\end{equation}

for a test particle falling to the brane from the exterior and
$-\Delta(\dot{r}^2)$ for a test particle exiting the bubble.

Let us consider now, as an example, a test particle which is
moving in a radial direction (i.e. $l = 0$) in the vicinity of a
static brane with a positive surface energy density. The spacetime
is chosen as de-Sitter in the interior ($A_{-}(r) = 1 -
\chi^2r^2$) and Schwarzschild in the exterior ($A_{+}(r) = 1 -
2GM/r$). In addition, we have $0<\lambda^2<1$, which gives $M>4\pi
R^3\rho_{0}/3$ (using $\chi^2 = 8\pi G\rho_{0}/3$).

Such a test particle, will experience a negative acceleration in
the exterior, due to the attractive characteristics of the
Schwarzschild spacetime, and a positive acceleration in the
interior, due to the repulsive nature of the de-Sitter spacetime.
This implies that test particles that lose energy in the vicinity
of the brane (e.g. by gravitational radiation and in particular
due to the gravitational radiation caused by the infinite
acceleration at the brane) will be trapped around the domain wall,
as it is evident from the form of the effective potential as
depicted in Fig. 1. While crossing the brane, the test particle
will experience an infinite acceleration whose sign is determined
by equation (33). We see that $\Delta(\dot{r}^2)> 0$ when the test
particle is entering the bubble and the opposite when it exists
the bubble. Thus, the infinite acceleration at the brane is
pulling the test particle towards the region with the lower
effective mass. The latter is a general feature of the theory and
does not depend on the particular chosen model.

Notice that in Newtonian theory there is no possibility for a
discontinuous velocity, since the Newtonian potential is always
continuous (in Newtonian theory one can allow only for a
discontinuity in the gradient of the potential) and there is no
sense in Newtonian theory for having a discontinuous mass. In our
theory we have defined a continuous potential, but had to pay by
having an effective, discontinuous mass. To show that the limit to
our theory indeed gives the correct non-relativistic physics we
take weak fields limit. This limit is describing weak
gravitational fields of static sources. Imposing this limit on
condition (33) gives us

\begin{equation}
    \lambda^2 = \frac{1-\frac{2GM}{R}}{1-\chi^2R^2} \approx
    \left(1 - \frac{2GM}{R}\right)\left(1 + \chi^2R^2\right)
    \approx 1 - \frac{2GM}{R} + \chi^2R^2,
\end{equation}

so that now (for finite energies and thus finite $\dot{r}$)

\begin{equation}
    \Delta(\dot{r}^2) =  v^2\left(\frac{1 - \lambda^2}{\lambda^2}\right) \approx v^2\left(\frac{\frac{2GM}{R} - \chi^2R^2}{1 - \frac{2GM}{R} + \chi^2R^2}\right) \approx
    0.
\end{equation}

Therefore, in the non-relativistic limit this effect disappears.

\section{Concluding Remarks and Discussion}
\label{ConcludingRemarksandDiscussion}

In this paper, we have shown that test particles can be trapped by
a static domain wall. As mentioned earlier, for the case of a stable
false vacuum bubble, there exist a variety of models. These models
use additional matter terms in the surface energy density function
of the domain wall, terms which are inserted "by hand". The
trapping of matter around the brane could give a physical
explanation for the origin of these terms, or give rise to a new
model in which the false vacuum bubble interacts with its physical
surroundings to form a stable soliton. It is interesting to note
that in the case of an anti-de Sitter bubble there is trapping
both in the center of the bubble as well as on the brane itself.

Moreover, we have found that there is a discontinuity in the
velocity of a test particle at the brane. This is a completely
unexpected phenomenon that can not be explained by the Newtonian
theory of gravity, and can only be studied correctly by using the
general theory of relativity.

Trapping by a negative surface tension was already considered in
\cite{bulg} in order to trap particles near the horizon of a black
hole. Now, we see that this effect is much more general and exists
for $\sigma > 0$ as well, and not necessarily around a horizon.

In the future, we wish to extend our study for the case of a
dynamic domain wall (i.e. $\dot{R} \neq 0$), as well as for the
study of the possibility for trapping of scalar and fermionic
fields around the domain wall of the bubble. The confinement of
scalar fields can lead to a physical explanation for the existence
of scalar fields on the brane, while trapping of fermionic fields
can explain, for example, the physical origin of the
Klinkhamer-Volovik model \cite{stable}.

We also expect large amount of all types of radiation when massive
particles will cross the brane, due to the large acceleration that
takes place in the process.

\vskip.3in

\centerline{{\bf Acknowledgments}} We wish to thank A. Davidson, I.
Gurwich, S. Rubin, A. Kaganovich, E. Nissimov, S. Pacheva and E.
Shafran for helpful discussions.

\vskip1.6in

\begin{figure}[htp]
\begin{center}
\includegraphics[scale=0.53]{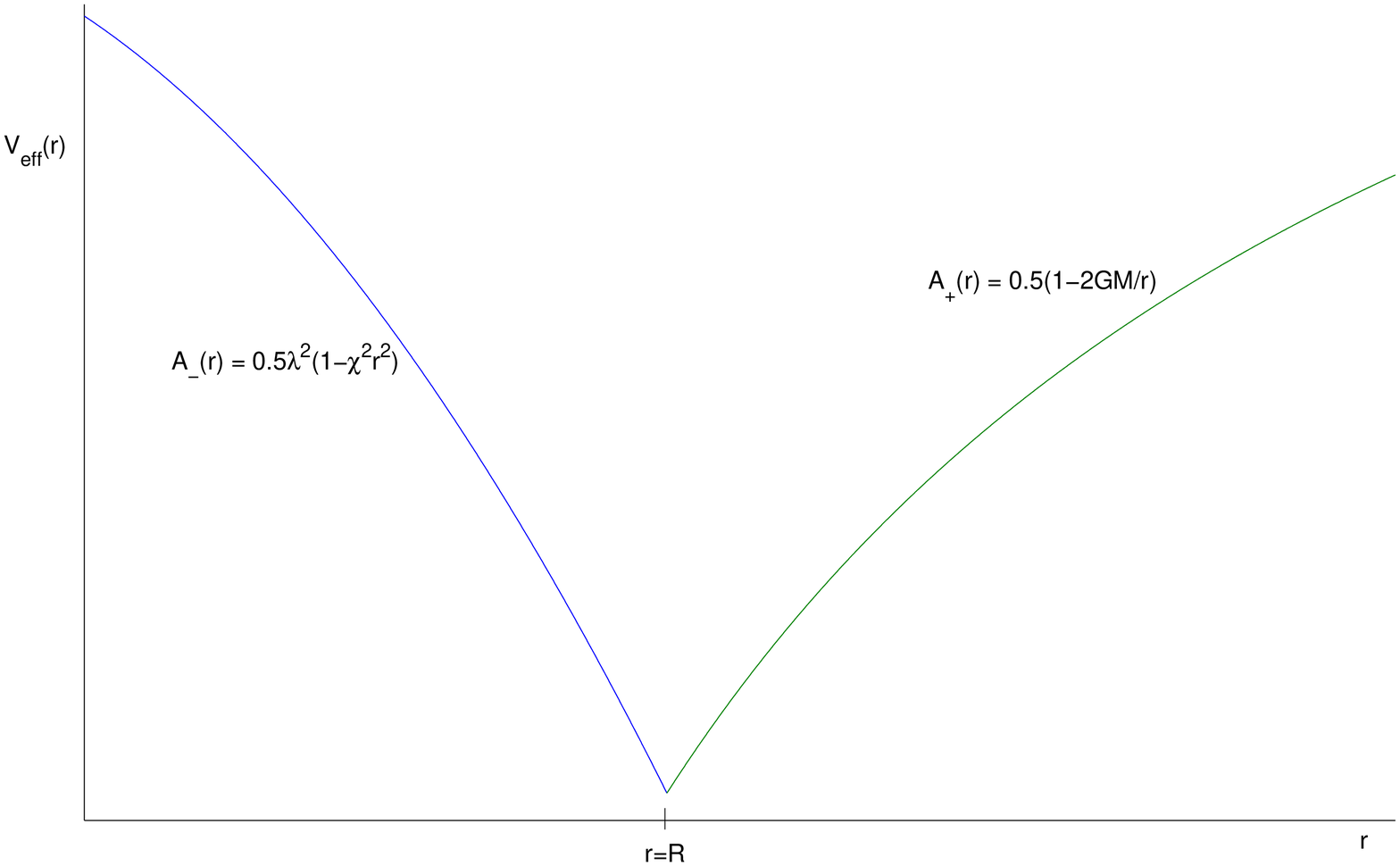}
\end{center}
\caption{An illustrative graph of the generic behavior of the
effective potential near the domain wall, for a radially moving
test particle in the vicinity of a static brane which separates a
de-Sitter region from a Schwarzschild region.} \label{fig:figure1}
\end{figure}

\end{document}